\documentclass[prb,aps,twocolumn,amsmath,amssymb,floatfix,showpacs,
superscriptaddress]{revtex4}

\usepackage[dvips]{graphics}
\usepackage{color}
\definecolor{gold}{rgb}{0.85,0.66,0}
\definecolor{dgreen}{rgb}{0.3,0,0.4}
\definecolor{dred}{rgb}{0.4,0,0.2}

\begin{document}

\title{\textcolor{dred}{Magneto-transport in a binary alloy ring}}

\author{Paramita Dutta}

\email{paramita.dutta@saha.ac.in}

\affiliation{Theoretical Condensed Matter Physics Division, Saha
Institute of Nuclear Physics, Sector-I, Block-AF, Bidhannagar,
Kolkata-700 064, India}

\author{Santanu K. Maiti}

\email{santanu@post.tau.ac.il}

\affiliation{School of Chemistry, Tel Aviv University, Ramat-Aviv,
Tel Aviv-69978, Israel}

\author{S. N. Karmakar}

\email{sachindranath.karmakar@saha.ac.in}

\affiliation{Theoretical Condensed Matter Physics Division, Saha
Institute of Nuclear Physics, Sector-I, Block-AF, Bidhannagar,
Kolkata-700 064, India}

\begin{abstract}
Magneto-transport properties are investigated in a binary alloy ring 
subjected to an Aharonov-Bohm (AB) flux $\phi$ within a single-band
non-interacting tight-binding framework. In the first part, we expose
analytically the behavior of persistent current in an isolated ordere 
binary alloy ring as functions of electron concentration $N_e$ and AB
flux $\phi$. While, in the second part of the article, we discuss 
electron transport properties through a binary alloy ring attached to 
two semi-infinite one-dimensional metallic electrodes. The effect of
impurities is also analyzed. From our study we propose that under suitable 
choices of the parameter values the system can act as a $p$-type or an 
$n$-type semiconductor.    
\end{abstract}



\maketitle

\section{Introduction}

The enormous advancements of nano-science and technology over the last few
decades have allowed us to investigate electron transport in low-dimensional
model quantum systems in a very tunable way. The precision instruments 
bestowed by nanotechnology have given indulgence to such low-dimensional 
systems with tailor-made geometries to become promising candidates for 
different nano-electronic devices. One of such geometries is a mesoscopic 
ring which looks very simple but has very high potential from the application 
perspective. Several quantum mechanical phenomena have been observed in this 
closed loop structure specially in presence of magnetic 
flux~\cite{butt1,buks,arunavada1,arunavada2,we}. The existence of 
dissipationless current, the so-called persistent current, in a mesoscopic 
normal metal ring pierced by an AB flux $\phi$ is one of such remarkable 
effects which reveals the importance of phase coherence of electronic wave 
functions in low-dimensional systems. In 1983, B\"{u}ttiker 
{\it et al.}~\cite{butt2} initially predicted this significant phenomenon 
through some theoretical arguments, but, its actual realization came much 
later after the nice experiment done by Levy {\it et al.}~\cite{levy}. 
They have observed the oscillations with period $\phi_0/2$ ($\phi_0=ch/e$,
the elementary flux-quantum) while measuring persistent current in an 
ensemble of $10^7$ independent Cu rings. Similar $\phi_0/2$ oscillations 
were also reported not only for an ensemble of disconnected $10^5$ Ag 
rings~\cite{mailly1} but also for an array of $10^5$ isolated GaAs-AlGaAs 
rings~\cite{mailly2,gefen}. Later many other theoretical~\cite{we2,san1,san2}
as well as experimental~\cite{mailly,blu} attempts have also been made 
to reveal the actual mechanisms of persistent current in single-channel
rings and multi-channel cylinders. Similar to persistent current in such
isolated closed geometries, the non-decaying charge current is also 
observed in open systems~\cite{butt3,akkermans,xia,mello,orellana,jaya,
xiong,pareek,bel1} like, a mesoscopic ring with side attached electrodes. 

Though a wealth of literature has already been generated involving the
analysis of persistent current in conventional mesoscopic rings, cylinders,
connected rings, ring with multi-arm structures, there is still need to 
look deeper into the problem to address several important issues 
those have not been explored earlier. In the present article we 
undertake an in-depth study of magneto-transport through a binary 
alloy mesoscopic ring in presence of the magnetic flux $\phi$ within
a single-band non-interacting tight-binding framework. We analyze our
results in two parts. In the first part, we describe the energy band 
structure and persistent current for an ordered binary alloy ring.
We perform these results analytically. In the second part, we describe
the magneto-transport properties in presence of external electrodes
\begin{figure}[ht]
{\centering \resizebox*{5.5cm}{4.2cm}{\includegraphics{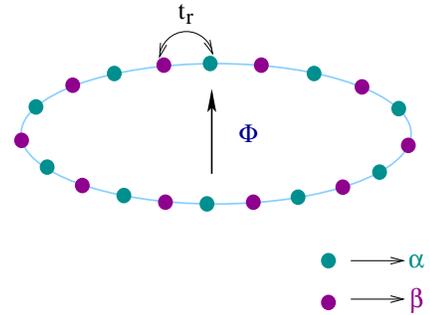}}\par}
\caption{(Color online). A binary alloy ring, threaded by a magnetic flux
$\phi$, is composed of two different types of atomic sites, viz, $\alpha$
and $\beta$ those are represented by filled green and magenta circles,
respectively.}
\label{ring1}
\end{figure}
which include transmission probabilities, persistent currents and related 
issues. Quite interestingly we examine that if we include some new additional 
atomic sites, called as impurities, those are not necessarily to be random, 
in any part of the binary alloy ring, some quasi-localized energy levels 
appear depending on the number of impurity sites within the band of extended 
regions. The positions of these nearly-localized energy levels can also be 
tuned with respect to the extended regions by means of some external gate 
voltages. It leads to the possibility of using such a system as a $p$-type 
or an $n$-type semiconductor by fixing the Fermi level at appropriate energy. 

In what follows, we describe the model quantum systems and the results. 
In Section II we present the results and related discussions for an
ordered binary alloy ring. Section III illustrates the model and the
theoretical approach to calculate transmission probability, persistent
current and related issues in presence of external leads, and the 
numerical results are described in Section IV. Finally, in Section V 
we draw our conclusions.

\section{Ordered binary alloy ring without external leads}

{\bf \underline{Model}:} The model quantum system is illustrated
schematically in Fig.~\ref{ring1}. An ordered binary alloy ring, subjected 
to an AB flux $\phi$, 
composed of two different types of atoms, namely, $\alpha$ and $\beta$ 
those are placed alternately in a regular pattern. We use a non-interacting
single-band tight-binding (TB) Hamiltonian to describe the model quantum 
of $\alpha$ and $\beta$ sites) binary alloy ring the TB Hamiltonian reads,
\begin{equation}
H_{R}=\sum_l (\epsilon_l c^{\dag}_l c_l + t_r e^{i \theta} c^{\dag}_l 
c_{l+1} + t_r e^{-i \theta} c^{\dag}_{l+1} c_{l})
\label{hr}
\end{equation}
where, $\epsilon_l$ is the on-site energy and $t_r$ refers to the 
nearest-neighbor hopping strength. For the site $\alpha$, 
$\epsilon_l$ is identical to $\epsilon_{\alpha}$, while it is 
$\epsilon_{\beta}$ for the site $\beta$. The phase factor $\theta=2\pi\phi/N$
appears into the Hamiltonian due to the magnetic flux $\phi$ threaded by
the ring which is measured by the elementary flux-quantum $\phi_0$.
$c_l^{\dagger}$ ($c_l$) is the creation (annihilation) operator for an
electron at the site $l$.

\vskip 0.2cm
\noindent
{\bf \underline{Energy spectrum}:} In order to explore the characteristic
features of persistent current in an ordered binary alloy ring, let us begin 
with the energy band structure. We do it analytically. The energy dispersion
\begin{figure}[ht]
{\centering \resizebox*{6.5cm}{4.6cm}{\includegraphics{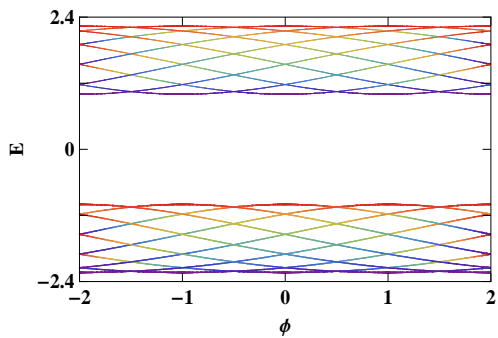}}\par}
\caption{(Color online). Energy-flux characteristics of an ordered binary
alloy ring ($N=40$) considering $\epsilon_{\alpha}=-\epsilon_{\beta}=1$
and $t_r=1$. $\phi_0$ is set at $1$.}
\label{band}
\end{figure}
relation for the ordered binary alloy ring can be expressed mathematically as,
\begin{equation}
E=\frac{\epsilon_{\alpha}+\epsilon_{\beta}}{2} \pm 
\sqrt{\left(\frac{\epsilon_{\alpha}-\epsilon_{\beta}}{2}\right)^2+
4\, t_r^2 \, cos^2{\left(ka\right)}}
\label{equ2}
\end{equation}
where, $a$ is the lattice spacing and $k$ is the wave vector. In presence
of the AB flux $\phi$, the quantized values of $k$ are obtained from the
relation, 
\begin{equation}
k=\frac{2 \pi}{N a} \left(n+\frac{\phi}{\phi_0}\right)
\end{equation}
where, $n$ is an integer and it is restricted within the range: $-N/2 \le 
n< N/2$. As illustrative example, in Fig.~\ref{band} we plot the energy 
levels as a function of flux $\phi$, obtained from Eq.~\ref{equ2}, for a 
$40$-site binary alloy ring considering 
$\epsilon_{\alpha}=-\epsilon_{\beta}=1$ and $t_r=1$. It shows that two 
different sets of energy levels are obtained, forming two quasi-bands, and 
they are separated by a finite energy gap. This gap, on the other hand, is 
tunable by the parameter values describing the TB Hamiltonian Eq.~\ref{hr}. 
The origin of two different sets of energy levels is also clearly visible
from Eq.~\ref{equ2}. From the energy spectrum (Fig.~\ref{band}) we see 
that at half-integer or integer multiples of flux-quantum, energy levels
have either a maximum or a minimum and it results vanishing nature of 
persistent current at these specific values of $\phi$, since the current
is obtained by taking the first order derivative of energy $E(\phi)$ with 
\begin{figure}[ht]
{\centering \resizebox*{7.5cm}{7cm}{\includegraphics{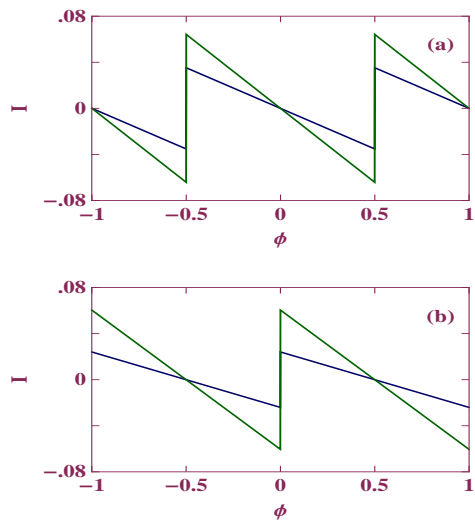}}\par}
\caption{(Color online). Current-flux characteristics of an ordered binary
alloy ring ($N=120$) considering $\epsilon_{\alpha}=-\epsilon_{\beta}=1$
and $t_r=1$. The blue and green colors in (a) correspond to $N_e=15$
and $35$, while in (b) they correspond to $N_e=10$ and $30$, respectively.
The lattice spacing $a$ is set at $1$ and we choose $\phi_0=1$.}
\label{oddeven}
\end{figure}
respect to flux $\phi$~\cite{san1}. All these energy levels vary 
periodically providing $\phi_0$ flux-quantum periodicity.

\vskip 0.2cm
\noindent
{\bf \underline{Persistent current}:} Once we get the energy eigenvalues
as a function of flux $\phi$, we can easily calculate persistent current
for individual energy eigenstates. It is simply the first order derivative 
of energy with respect to flux $\phi$. Therefore, for an $n$-th energy
eigenstate we can write the expression for the current as,
\begin{equation}
I_n=\pm \left(\frac{4 \pi t_r^2}{N a \phi_0} \right)
\frac{sin{\left[\frac{4 \pi}{N a}\left(n+\phi/\phi_0\right)\right]}}
{\sqrt{1+ 4 t_r^2 cos^2{\left[ \frac{2 \pi}{N a} \left(n+\phi/\phi_0\right)
\right]}}}
\end{equation}
where, $+$ve or $-$ve sign in the current expression appears depending on the
choice of $n$ i.e., in which sub-band the energy level exists. At absolute
zero temperature, total persistent current $I$ for a particular filling $N_e$
is obtained by taking the sum of individual contributions from the lowest
$N_e$ energy eigenstates. Thus, we can write, 
\begin{equation}
I=\sum_{n=1}^{N_e} I_n
\end{equation}
As representative examples, in Fig.~\ref{oddeven} we display the variation
of persistent current for an ordered $120$-site binary alloy ring considering
$\epsilon_{\alpha}=-\epsilon_{\beta}=1$ and $t_r=1$ in the different filled
band cases. For odd number of electrons the results are shown in 
Fig.~\ref{oddeven}(a) while, in Fig.~\ref{oddeven}(b) the results are given
for even number of electrons. Both for the odd and even $N_e$, persistent
current shows saw-tooth like variation as a function of flux $\phi$, similar 
to that of traditional single-channel mesoscopic rings. The sharp transitions
at half-integer (for odd $N_e$) or integer (for even $N_e$) multiples of
flux-quantum ($\phi_0$) in persistent current appears due to the crossing
of energy levels at these respective values of $\phi$. Quite interestingly,
we also examine that the current shows always diamagnetic in nature 
irrespective of the filling factor. 

\section{Binary alloy ring with side-attached leads}

In the rest of the present article we explore the essential features of
magneto-transport properties for a binary alloy ring in presence of 
external leads i.e., for an open system.

\begin{figure}[ht]
{\centering \resizebox*{8.0cm}{4.8cm}{\includegraphics{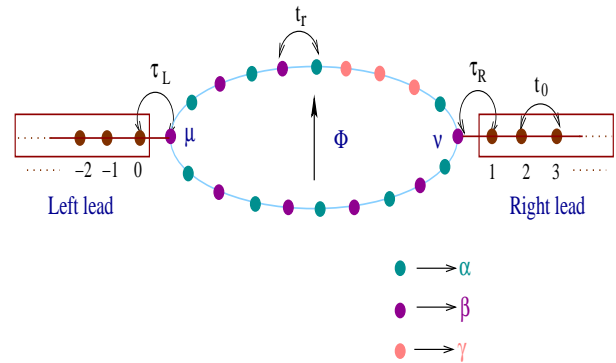}}\par}
\caption{(Color online). A binary alloy ring in presence of some identical
foreign atoms (labeled as $\gamma$ sites), threaded by an AB flux $\phi$,
is attached to two semi-infinite one-dimensional metallic leads. $\mu$ and
$\nu$ are the connecting sites.}
\label{ring2}
\end{figure}

\vskip 0.2cm
\noindent
{\bf \underline{Model}:} The model quantum system is depicted in
Fig.~\ref{ring2} where a binary alloy ring, threaded by a magnetic flux 
$\phi$, is attached to two semi-infinite one-dimensional metallic electrodes.
The ring is composed of $N_1$ identical pairs of $\alpha$-$\beta$ sites
and $N_2$ identical foreign atomic sites, labeled as $\gamma$ sites, those 
are embedded in a small portion of the ring. These sites ($\gamma$ sites) 
are often called as impurity sites. We label the atomic sites of two 
side-attached leads in a particular way, as shown in Fig.~\ref{ring2},
and they are connected to the ring via the sites $\mu$ and $\nu$. A
single-band non-interacting tight-binding framework is used to describe
the entire system. For the full system we can write the total Hamiltonian
as a sum of three terms like,
\begin{equation}
H=H_{R}+H_L+H_T
\label{part}
\end{equation}
where, $H_R$, $H_L$ and $H_T$ correspond to the Hamiltonians for the 
ring, leads (left and right) and tunneling between the ring and leads, 
respectively. The first term ($H_R$) looks exactly similar as in 
Eq.~\ref{hr}, but here $\epsilon_l$ has three possibilities for three
different atomic sites ($\alpha$, $\beta$ and $\gamma$). We can also
tune the site energy $\epsilon_{\gamma}$ by means of some external gate
voltage $V_g$, and thus, we can write it as $\epsilon_{\gamma}=
\epsilon_{\gamma}^0 + V_g$, where $\epsilon_{\gamma}^0$ is the site
energy in absence of any external potential. The other two terms of
Eq.~\ref{part}, $H_L$ and $H_T$, can also be expressed in a similar 
fashion as,
\begin{equation}  
H_L=\underbrace {t_0 \sum_{m \le 0} \left(b_m^{\dag} b_{m-1} + 
h.c. \right)}_{\mbox {left lead}} +
\underbrace {t_0 \sum_{m \ge 1} \left(b_m^{\dag}b_{m+1} +
h.c. \right)}_{\mbox {right lead}}
\label{hlead}
\end{equation}
and,
\begin{equation}
H_T=\left( \tau_L b_0^{\dag} c_{\mu} + \tau_R b_1^{\dag} c_{\nu} \right) +
h.c.
\end{equation}
The site energy dependent term in $H_L$ is omitted as we set the site
energy for the identical sites in the leads to zero. $b_m^{\dag}$ ($b_m$) 
corresponds to the creation (annihilation) operator of an electron at the
site $m$ of the leads and $t_0$ represents the nearest-neighbor hopping
integral in these leads. $\tau_L$ is the coupling strength between the
left lead and the ring, while it is $\tau_R$ for the other case.

\vskip 0.2cm
\noindent
{\bf \underline {Wave-guide theory}:}
To find transmission probability across the ring and also to calculate
persistent current in such an open ring system we adopt wave-guide 
theory~\cite{xia,jaya,xiong}. Here we describe it very briefly.

Let us start with the Scr\"{o}dinger equation $H |\psi \rangle = E |\psi 
\rangle$, where $|\psi \rangle$ is the stationary wave function of the 
entire system. In the Wannier basis it ($|\psi \rangle$) can be written as,
\begin{equation}
|\psi \rangle= \underbrace{\sum_{m \le 0} B_m |m \rangle}_{\mbox{left lead}} 
+ \underbrace{\sum_{m \ge 1} B_m |m \rangle}_{\mbox{right lead}} + 
\underbrace{\sum_{l} C_l |l \rangle}_{\mbox{ring}}
\label{shy}
\end{equation}
where, the co-efficients $B_m$ and $C_l$ correspond to the probability
amplitudes in the respective sites. Assuming that the electrons are described
by a plane wave we can express the wave amplitudes in the left and right
leads as,
\begin{equation}
B_m=e^{ikm}+r e^{-ikm}, \hskip 0.3cm {\mbox{for}} \hskip 0.3cm m \le 0
\label{equ10}
\end{equation}
and
\begin{equation}
B_m=t e^{ikm}, \hskip 0.3cm {\mbox{for}} \hskip 0.3cm m \ge 1 
\label{equ11}
\end{equation}
where, $r$ and $t$ are the reflection and transmission amplitudes, 
respectively. $k$ is the wave number and it is related to the energy 
$E$ of the incident electron by the expression $E=2t_0 \cos{k}$. The
lattice spacing $a$ is set equal to $1$. 

Now to find out the transmission amplitude $t$, we have to solve the 
following set of coupled linear equations.
\begin{eqnarray} 
E B_0 &=& t_0 B_{-1} + \tau_L C_{\mu} \nonumber \\
(E-\epsilon_l) C_l &=&t _r e^{i \phi} C_{l+1}+t_r e^{-i \phi} C_{l-1}+
\tau_L B_0 \delta_{l,\mu} \nonumber \\
~& &~~~~~~~~~~~~+ \tau_R B_1 \delta_{l,\nu} \nonumber \\
E B_1&= &t_0 B_2 + \tau_R C_{\nu} 
\label{seteqns}
\end{eqnarray}
where, the co-efficients $B_0$, $B_{-1}$, $B_1$ and $B_2$ can be easily 
expressed in terms of $r$ and $t$ by using Eqs.~\ref{equ10} and \ref{equ11},
and they are in the form:
\begin{eqnarray}
B_0 &=& 1+r \nonumber \\
B_{-1}&=& B_0 e^{ik}-2i \sin{k} \nonumber \\
B_1 &=& t e^{ik} \nonumber \\
B_2 &=& t e^{2ik}
\end{eqnarray}
Thus, for a particular value of $E$ we can easily solve the set of linear 
equations and find the value of $t$. Finally, the transmission probability
across the ring becomes 
\begin{equation}
T(E)=|t|^2
\end{equation}
Now, to calculate persistent current between any two neighboring sites 
in the ring we use the following relation,
\begin{equation} 
I_{l,l+1}=\frac{2 e t_r}{N \hbar} {\mbox{Im}}\left( C_l^{\dag} C_{l+1} 
e^{-i \phi}
\right).
\end{equation}

\section{Numerical Results and discussion}

Throughout the calculations we set $\epsilon_{\alpha}=-\epsilon_{\beta}=1$, 
$\epsilon_{\gamma}^0=0$, $t_r=1$, $\epsilon_0=0$ and $t_0=2$. The energy
scale is measured in unit of $t_r$ and we choose the units where $c=h=e=1$.

\subsection{Transmission and ADOS spectra}

\begin{figure}[ht]
{\centering \resizebox*{6.8cm}{10.6cm}{\includegraphics{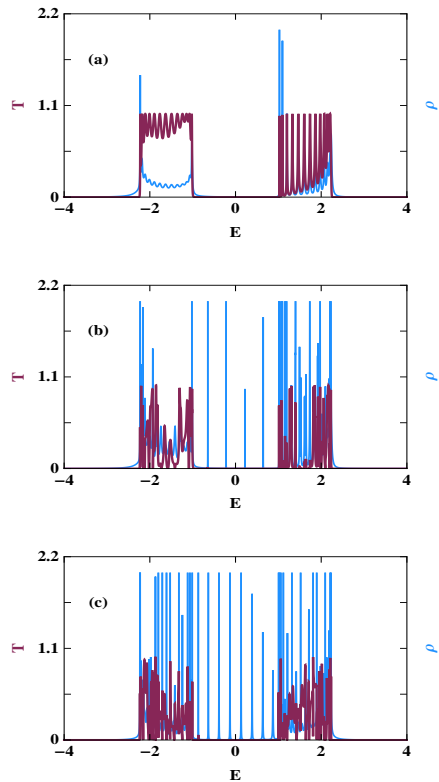}}\par}
\caption{(Color online). Transmission probability (red color) and average 
density of states (blue color) for some typical binary alloy rings with
fixed number of $\alpha$-$\beta$ pairs ($N_1=28$), but different values of
impurity sites $N_2$, where (a) $N_2=0$, (b) $N_2=12$ and (c) $N_2=22$.
For these three cases we set $\mu=1$ and choose $\nu=29$, $35$ and $40$,
respectively. Other parameters are: $\epsilon_{\gamma}=0$, $\tau_L=\tau_R=1$
and $\phi=0$.}
\label{spectrum}
\end{figure}

In Fig.~\ref{spectrum} we show the variation of transmission probability
$T$ (red color) as a function of injecting electron energy $E$ for some 
typical binary alloy rings considering different number of impurity sites.
The average density of states (ADOS) is also superimposed in each spectrum.
Here, we fix $N_1=28$ and take three different values of $N_2$ those are 
presented in Figs.~\ref{spectrum}(a), (b) and (c), respectively. In all 
these cases the magnetic flux $\phi$ is set equal to zero. The results are
quite interesting and important also. It is observed that in absence of
impurity sites electron can transmit through the ring for two wide band of
energies (see Fig.~\ref{spectrum}(a)). This transmission spectrum exactly 
overlaps with the ADOS profile ($\rho$-$E$ spectrum) which ensures that 
electronic transmission takes place through all the energy eigenstates 
of the binary alloy ring and they are extended in nature.
The band splitting for the ordered binary alloy ring is easily understood
from our previous discussion (Sec. II) and the gap between these two bands
are also tunable by the parameter values describing the system. The situation
becomes really interesting when some additional impurities ($\gamma$ sites)
are introduced in any part of the binary alloy ring. Due to the inclusion 
of such atomic sites some energy levels appear within the band of extended
regions those are no longer extended, but they are almost quasi-localized
and do not contribute to the electronic transmission. This behavior is well
explained in Figs.~\ref{spectrum}(b) and (c), where $T$-$E$ and $\rho$-$E$
spectra are superimposed with each other for two different numbers of 
impurity sites ($N_2$). The transmission of electrons takes place mainly
along the two band edges, while in the band centre, several energy levels
are there which provide zero transmission probability. The number of these 
almost localized energy levels between the extended regions gets increased
with the increase of impurity sites, and for large enough impurity sites
they almost form a quasi-energy band of localized states. The location of 
the almost localized energy band can be shifted towards the edge of extended
\begin{figure}[ht]
{\centering \resizebox*{7.5cm}{8.6cm}{\includegraphics{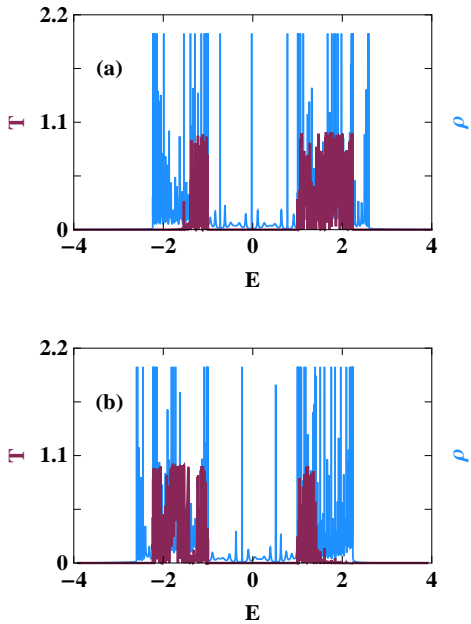}}\par}
\caption{(Color online). Transmission probability (red color) and ADOS 
(blue color) as a function of energy for a binary alloy ring ($N_1=60$) 
with $44$ impurity sites ($N_2=44$), where (a) and (b) correspond to
$\epsilon_{\gamma}=0.6$ and $-0.6$, respectively. Other parameters are:
$\phi=0$, $\tau_L=\tau_R=1.5$, $\mu=1$ and $\nu=83$.}
\label{pntype}
\end{figure}
regions simply by tuning the site energy of these foreign atoms, and here we
do it by means of applying an external gate voltage $V_g$. This phenomenon
can be utilized to use such a geometry as a $p$-type or an $n$-type 
semiconductor
by fixing the Fermi level in an appropriate place. To explore this fact in 
Fig.~\ref{pntype} we present the variation of transmission probability
in addition to the ADOS for a much larger ring size where we consider 
$60$ identical pairs of $\alpha$-$\beta$ sites and $44$ impurity sites.
Two different cases are considered, one for the case when $\epsilon_{\gamma}$
is set at $0.6$ (Fig.~\ref{pntype}(a)), and the other is shown for 
$\epsilon_{\gamma}=-0.6$ (Fig.~\ref{pntype}(b)). For a quite large number
of impurity sites we get almost quasi-energy bands (ADOS spectra) and quite
nicely we see that when $\epsilon_{\gamma}$ is fixed at $0.6$, a localized
energy band for a wide range of energy is formed along the left edge of 
the extended region (Fig.~\ref{pntype}(a)). Now if we choose the Fermi 
level around $E=-1.7$, then many electrons in the localized region below 
the Fermi level can jump, even at much low temperature since the energy gap
is almost zero, to the extended regions and can contribute to the current.
Thus, we get a large number of excess electrons, depending on the localized 
energy levels and also the available extended energy states, in the conduction
region which behave as n-type carriers. An exactly opposite behavior is
obtained when we set $\epsilon_{\gamma}$ to $-0.6$. In this case the wide
band of localized states is formed in the right edge of the extended region
\begin{figure}[ht]
{\centering \resizebox*{8cm}{7.6cm}{\includegraphics{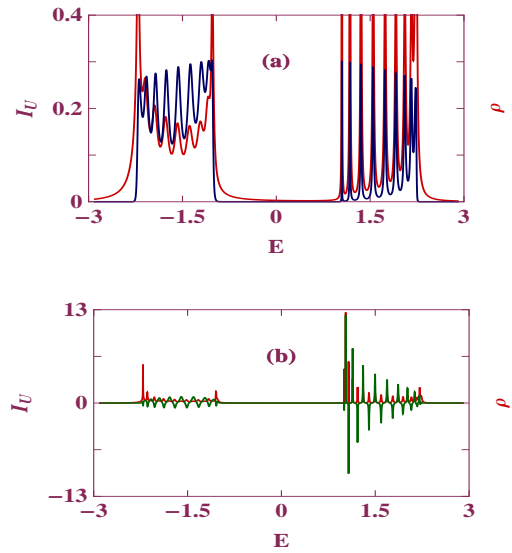}}\par}
\caption{(Color online). Persistent current ($I_U$) in the upper arm as
a function of energy $E$ for a binary alloy ring ($N_1=20$) in the absence
of impurity atoms ($N_2=0$), where (a) $\phi=0$ and (b) $\phi=\phi_0/4$.
Other parameters are: $\tau_L=\tau_R=0.8$, $\mu=1$ and $\nu=21$. For this
ordered ring, persistent current in the lower arm ($I_L$) is exactly
identical to $I_U$. ADOS profile (red color) is also superimposed in each
spectrum.}
\label{curr1}
\end{figure}
(see Fig.~\ref{pntype}(b)). Therefore, if we set the Fermi energy level 
around $E=1.7$, then the electrons from the filled extended levels below
the Fermi level hop to the nearly empty localized levels, and these electrons
do not contribute anything to the current. But some holes are formed in the
extended regions which can carry current and we get $p$-type carriers. So,
in short, we can emphasize that by fixing the Fermi level in appropriate 
places the geometry can be utilized as a $p$-type or an $n$-type 
semiconductor.

Before we end this analysis, we would like to point out
that here we present the results for a particular set of parameter values 
and establish how such a geometry can be utilized as a $p$-type or an 
$n$-type semiconductor upon the movement of the Fermi level. For a model 
calculation we choose some specific values of the parameters used in 
the numerical calculation, but all these physical phenomena are exactly 
invariant with the change of the parameter values. Only the numerical 
values will be altered. These features are also exactly valid even for 
a non-zero value of magnetic flux $\phi$. The physical picture will be 
much more appealing if we consider larger rings with more impurity sites
and it gives us the confidence to propose an experiment in this line.
In a recent work Bellucci {\em et al.}~\cite{bel1} have done a detailed
study of magneto-transport properties in quantum rings considering tunnel 
barriers in the presence of magnetic field and established how 
metal-to-insulator transition takes place in such a geometry. It has been 
shown that by controlling the strength and the positions of the barriers, 
the energy shift can be done in a tunable way and they have also proposed
this model for experimental realization. This is quite analogous to our 
present study, and therefore, an experiment in this regard will be
challenging.

\subsection{Persistent current}
 
The behavior persistent current $I_U$ as a function of energy $E$
in the upper arm of an ordered binary alloy ring is presented in 
Fig.~\ref{curr1}, where (a) and (b) correspond to $\phi=0$ and $\phi_0/4$,
respectively. The ring is symmetrically coupled to the side attached leads
i.e., the upper and lower arms have identical length, and accordingly, the 
\begin{figure}[ht]
{\centering \resizebox*{8cm}{7.6cm}{\includegraphics{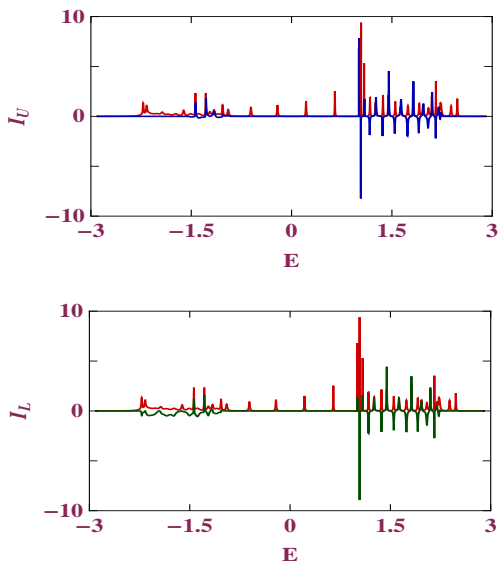}}\par}
\caption{(Color online). Persistent current as a function of energy for a
binary alloy ring ($N_1=16$) in presence of impurity sites ($N_2=12$) for
a finite value of $\phi$ ($\phi=0.3$), where (a) and (b) correspond to the
currents for the upper and lower arms respectively. Other parameters are:
$\epsilon_{\gamma}=0.5$, $\tau_L=\tau_R=0.8$, $\mu=1$ and $\nu=23$. ADOS
profile (red color) is superimposed in each spectrum.}
\label{curr2}
\end{figure}
current $I_L$ in the lower arm becomes exactly identical to the current 
obtained in the upper arm. From the spectra we observe that the current 
is available for two wide range of energies, separated by a finite gap, 
associated with the energy levels of the ring those are clearly visible 
from the ADOS profiles (red color). All these energy eigenstates are 
extended in nature since they provide a non-zero current, but the nature
of the current is quite different for the two different cases i.e., when
$\phi=0$ (Fig.~\ref{curr1}(a)) and $\phi \ne 0$ (Fig.~\ref{curr1}(b)). 
For $\phi=0$, persistent current shows $+$ve value for all the energy 
levels which predicts that the currents are in the same phase. While,
for a finite value of $\phi$, the phase of the current changes 
alternatively. So by measuring persistent current we can directly estimate 
the nature of the current i.e., whether it is paramagnetic or diamagnetic
in nature and also predict the characteristics of energy eigenstates, which 
are somewhat interesting in the study of electron transport.

To explore the effect of impurities on persistent current in such a geometry,
in Fig.~\ref{curr2}, we display the results for a typical binary alloy ring
($N_1=16$) in presence of some impurity sites ($N_2=12$), where (a) and (b)
represent the results for the upper and lower arms of the ring, respectively.
The impurity sites are added in the upper arm of the ring. With the
inclusion of impurity sites, the symmetry between the two arms is broken,
and therefore, the currents in the upper and lower arms are no longer
identical to each other as shown from the spectra (Figs.~\ref{curr2}(a) and
(b)). Unlike to the ordered binary alloy ring (Fig.~\ref{curr1}), here 
all the energy eigenstates are not extended in nature. Some localized
energy levels appear in the band of extended energy states due to the
presence of impurity sites in the ring. This is clearly visible from the
spectra since for these impurity levels no current is available. Thus,
be calculating persistent current we can emphasize the nature of energy
eigenstates very nicely, and this idea can be utilized to reveal the 
localization properties of energy eigenstates in any complicated geometry.

Analogous to our present model, few works~\cite{bel1,bel2} 
have also been done very recently in some mesoscopic rings in presence 
of tunnel barriers or in presence of an artificial crystal. The nature of
energy band structures and the behavior of persistent currents have been
described in detail in these samples, and, we hope our results provide some 
extensions in this regard.

\section{Conclusion}

In the present article we make a detailed study of magneto-transport
through a binary alloy ring in presence of magnetic flux $\phi$ within a
single-band non-interacting tight-binding framework. The essential results
are analyzed in two parts. In the first part, we discuss the band structure 
and persistent current of an isolated ordered binary alloy ring. These 
results are done analytically. In the other part, we explore the 
magneto-transport properties of a binary alloy ring in presence of 
external leads. The effect of impurities are also addressed. Quite
interestingly we see that in presence of some foreign atoms, those
are not necessarily be random, in any part of the ring, some quasi-localized
energy levels appear within the band of extended energy levels. The locations
of these almost localized energy levels can also be regulated by means of
some external gate voltage. This leads to a possibility of using such a 
system as a $p$-type or an $n$-type semiconductor by fixing the Fermi level
in appropriate places.


\begin{thebibliography}{99}

\bibitem{butt1} A. L. Yeyati and M. Buttiker, Phys. Rev. B \textbf{52}, 
R14360 (1995).

\bibitem{buks} E. Buks, R. Schuster, M. Heiblum, D. Mahalu, V. Umansky, 
and H. Shtrikman, Phys. Rev. Lett. \textbf{77}, 4664 (1996).

\bibitem{arunavada1} S. Jana and A. Chakrabarti, Phys. Rev. B \textbf{77}, 
155310 (2008). 

\bibitem{arunavada2} S. Jana and A. Chakrabarti, Phys. Satus Solidi (b)
\textbf{248}, 725 (2011). 

\bibitem{we} P. Dutta, S. K. Maiti, and S. N. Karmakar, Solid State Commun.
\textbf{150}, 1056 (2010).

\bibitem{butt2} M. B\"{u}ttiker, Y. Imry, and R. Landauer, Phys. Lett. A 
\textbf{96}, 365 (1983).

\bibitem{levy} L. P. Levy, G. Dolan, J. Dunsmuir, and H. Bouchiat, Phys.
Rev. Lett. \textbf{64}, 2074 (1990).

\bibitem{mailly1} R. Deblock, R. Bel, B. Reulet, H. Bouchiat, and D. Mailly,
Phys. Rev. Lett. \textbf{89}, 206803 (2002).

\bibitem{mailly2}  B. Reulet, M. Ramin, H. Bouchiat, and D. Mailly, Phys. 
Rev. Lett. \textbf{75}, 124 (1995).

\bibitem{gefen} H. F. Cheung, Y. Gefen, E. K. Reidel, and W. H. Shih, Phys. 
Rev. B \textbf{37}, 6050 (1988).

\bibitem{we2} P. Dutta, S. K. Maiti, and S. N. Karmakar, Eur. Phys. J. B
(in press).

\bibitem{san1} S. K. Maiti, M. Dey, S. Sil, A. Chakrabarti, and S. N.
Karmakar, Europhys. Lett. \textbf{95}, 57008 (2011).

\bibitem{san2} S. K. Maiti, Solid State Commun. \textbf{150}, 2212 (2010).

\bibitem{mailly} D. Mailly, C. Chapelier, and A. Benoit, Phys. Rev. Lett.
\textbf{70}, 2020 (1993).

\bibitem{blu} H. Bluhm, N. C. Koshnick, J. A. Bert, M. E. Huber, and
K. A. Moler, Phys. Rev. Lett. \textbf{102}, 136802 (2009).

\bibitem{butt3} M. B\"{u}ttiker, Phys. Rev. B \textbf{32}, R1846 (1985).

\bibitem{akkermans} E. Akkermans, A. Auerbach, J. E. Avron, and B. Shapiro,
Phys. Rev. Lett. \textbf{66}, 76 (1991).

\bibitem{xia} J.-B. Xia, Phys. Rev. B \textbf{45}, 3593 (1992).

\bibitem{mello} P. A. Mello, Phys. Rev. B \textbf{47}, 16358 (1993).

\bibitem{orellana} P. A. Orellana, M. L. Ladr\'{o}n de Guevra, M. Pacheco, 
and A. Latg\'{e}, Phys. Rev. B \textbf{68}, 195321 (2003).

\bibitem{jaya} A. M. Jayannavar and P. Singha Deo, Phys. Rev B \textbf{49},
13685 (1994).

\bibitem{xiong} Y.-J. Xiong and X.-T. Liang, Phys. Lett. A \textbf{330},
307 (2004).

\bibitem{pareek} T. P. Pareek, P. Singha Deo, and A. M. Jayannavar, Phys. 
Rev. b \textbf{52}, 14657 (1995).

\bibitem{bel1} S. Bellucci and P. Onorato, Physica E \textbf{41}, 1393 (2009).

\bibitem{bel2} S. Bellucci and P. Onorato, Eur. Phys. J. B \textbf{73}, 215 
(2010).


\end{thebibliography}
\end{document}